# Experimental study of photon beam polarimeter based on nuclear $e^+ e^-$ pair production in an amorphous target

F.Adamyan, A.Aganyants, H.Hakobyan, J.Manukyan,
R.Oganezov, L.Sargsyan, A.Sirunyan[*)] and H.Vartapetian

Yerevan Physics Institute, 2 Alikhanian Brothers str., EPD, 375036, Yerevan, Armenia

R.Jones

University of Connecticut, Storrs, CT, USA

## Abstract

The experimental method of the linearly polarized photons polarimetry, using incoherent $e^+ e^-$ - pair production process has been investigated on the beam of coherent bremsstrahlung (CB) photons in the energy range of 0.9 – 1.1 GeV at the Yerevan synchrotron.

## Introduction

Direct methods for linearly polarized (CB) photons polarimetry is possible to realize by measuring azimuthal asymmetry of $e^+ e^-$ - pairs photoproduction in an oriental crystal [1], in an amorphous target [2] or atomic electrons [3]. The precise CB polarimetry may also be realized by means of calculational methods were based on the measured CB intensity spectrum and shape analysis [4,5].

In the recent article [6], description of CB polarimeter for directly measuring the linear polarization a photon beam is presented, exploiting an azimuthal dependence of incoherent $e^+ e^-$ pair photoproduction on a nuclei within a narrow ranges in both of polar $\Delta\theta$ and azimuthal $\Delta\varphi$ angles [7]. The Monte Carlo simulations have shown a feasibility of polarimetry precision in the level of $\sigma_p = 0.02$ at CB energy range $E_\gamma = 0.9\text{-}1.1$ GeV if symmetric $e^+e^-$ pairs are selected.

In this paper we are presenting and discussing the results of polarimeter's experimental study, which has been recently constructed and installed in a γ-2 beam line of YERPHI's electron synchrotron, also experimental results of the first measurements on CB photon beam with maximal energy $E_\gamma = 2,55$ GeV are presented.

## 1. Method of polarization measurement

The analyzing power or azimuthal asymmetry of incoherent e-e+ pairs production process is defined as:

$$A = \frac{\sigma_{\text{II}} - \sigma_{\perp}}{\sigma_{\text{II}} + \sigma_{\perp}}, \tag{1}$$

---

[*)]Corresponding author, Tel.: +3741 342747, E-mail: sirunian@mail.yerphi.am (A.Sirunyan)

where $\sigma_\parallel$, $\sigma_\perp$ are the differential cross-sections for the azimuthal angle of pair production plane parallel, perpendicular to the plane of photon polarization.

For the cross-sections calculations, an analytical expressions of ref. [8] are used, where the degree of linear polarization is described in the terms of Stokes parameter $\xi_3$. The values of $\xi_3 = +1, -1$ correspond to 100% of linear polarization ($P_\gamma = 1$) both for the polarization orientation perpendicular and parallel to the production plane.

The CB photon beam linear polarization ($P_\gamma < 1$) is related to a measured asymmetry by

$$P_\gamma = A_{exp} / A_{MC} \qquad (2)$$

where $A_{exp}$ is the experimental asymmetry and $A_{MC}$ is the Monte-Carlo simulation result, calculated for $P_\gamma = 1$. The simulation includes modeling pair production using differential cross-sections with atomic form-factors [6,9], all experimental conditions and details of polarimeter PS-6. An application of the expression (2) assumes an equality of CB's intensity and polarization spectra. The precision of $P_\gamma$ determination depends on the statistic and systematic uncertainties in the values of $A_{exp}$ and $A_{MC}$, so the full simulation is really needed, aimed to define and hold under control all sources of the systematic uncertainties.

## 2. Layout of the experimental setup

The sketch of experimental setup arrangement in a γ-2 beam line is presented in Fig.1. The beam of linearly polarized photons [10], generated by 2.55 GeV electrons on a diamond crystal (length 8 mm, width 2 mm thickness 0,072 mm), is collimated and cleaned by the set of collimators $K_1$, $K_2$ and sweeping magnets $SM_1$, $SM_2$ to a angular divergence of the beam to a half-angle of 0.17 mr and passes through 10μm of thin Mylar converter ($C_1$) located at the entrance of PS-30 pair spectrometer. The pair spectrometer allows to measure a CB intensity spectrum simultaneously in the thirty energies points with energy resolution $\delta E_\gamma / E_\gamma = 0.02$ [11]. An integral intensity of the photon beam is measured by Wilson quantameter (Q). The polarimeter PS-6 includes a 20 μm thickness aluminum converter ($C_2$), vertical slit collimator ($K_3$), which provides angular selection of the emitted $e^+ e^-$ pairs, and horizontally bending magnet equipped with a six telescopes of the scintillating counters, three hodoscope elements ($N_1 \div N_3$) in each arm (Fig.2). The vertical slit, installed in the vacuum pipe, has been made of 6 cm thickness lead and has a 26 mm width, 80 mm height, while γ-beam profile has a 16x16 mm$^2$. The telescopes are formed by the coincidence of three small forward counters and one big backward counter. The left and right telescopes are shifted up and down relative to a median plane of PS-6 spectrometer, allowing to select the symmetric $e^+ e^-$ pairs. The PS-6 hodoscopes are located 19.9 m downstream of the $C_2$ converter and select polar angles range. The full beam line between the collimator $K_2$ and the exit of PS-6 is vacuum pumped, allowing to decrease the multiple scattering influence. The hodoscopes counters widths of $N_1 \div N_3$ are chosen to be 2.5, 5.0 and 2.5 cm, that corresponds to a 3 independent energy bins with energy resolution of $\sigma_E$ = 12, 20 and 12 MeV for symmetric combination of registered pairs at CB peak setting to 1000 MeV. The vertical movement of the telescopes is performed distantly with precision of 0.1mm.

## 3. Monte-Carlo simulations and polarimeter alignment

The detailed presentation of Monte Carlo calculations is done in ref. [6]. Here we are considering and discussing an influence of some experimental uncertainties to the polarimeter's analyzing power. The precision of the polarimeter geometry in a sense is quite important, involving in particular a precision in positioning of the telescopes in the vertical (z) and horizontal (x,y) directions also alignment of the vertical slit relative the photons' beam. Above mentioned uncertainties may become a source of the systematic errors if the geometry of Monte Carlo simulations differs from the actual one. In this respect a number of calculations has been done to define an expected dependences and elaborate a necessary experimental tests for the geometry control.

The Fig.3 shows dependence of the analyzing power $A_{MC}$, calculated for Cromer-Waber form factor [6] and for energy of the peak in CB photon spectrum $E_\gamma^{peak}$ = 1.0 GeV from the vertical shift ($\Delta z$) of the central telescopes ($N_{2L}$, $N_{2R}$) relative the median plane with and without the vertical-slit installed. The experimental study of $A_{MC}$'s z dependence and its annulling within the limits of the statistical and systematical uncertainties in the vicinity of $z \approx 0$, is an important test of the polarimeter performance. As can be seen in the Fig. 3, the most optimal value for z selection is the low gradient flat zone around z=10 mm. An uncertainty in the z setting may mainly arise due to the uncertainty of the median plane position at the location of telescopes.

A test is proposed to determine the position of the median plane. At the fixed vertical gap between the up and down telescopes, the z-scan around expected position of the median plane is performed (see Section 4). Fig.4 shows a Monte-Carlo calculation results obtained for configuration of PS-6 z=0.

An influence of x-coordinate uncertainty of the beam position in the median plane has been investigated. The shift of the telescopes toward the x-direction leads to $A_{MC}$ asymmetry changes as it is shown in Fig.5 for the central telescopes ($N_{2up}$ x $N_{2down}$). This graph is made for a case when $N_{2up}$ is fixed to the x=0 position, and $N_{2down}$ will shift to right on $\Delta x$.

For the experimental check of PS-6 detectors positioning symmetry around vertical and horizontal axes (z, x) is necessary to measure and compare the coincidence rates for the detectors configurations presented in Fig.6. The Monte Carlo calculations for these configurations gave a compatible results within the limits of statistical uncertainties. Symmetry of detectors together with polarimeter`s magnetic field needs to check experimentally at these two configurations (rotations around vertical and horizontal axes).

An application of all the proposed tests allows to determine the actual geometry of the polarimeter and evaluate an expected systematic errors in the values of measured asymmetry.

**4. Results of the first meaurements**

The measurements have been carried out on the beam of linearly polarized photons with intensity of $10^8 \gamma/s$ at the CB peak energy setting to $E_\gamma$=1000 MeV. The spectrum in the peak region was measured and monitored each 3 minutes and peak position controlled. In the case of the peak energy shift above $\Delta E_\gamma/E_\gamma$=0.02 due to beam instability, the data taking was blocked and crystal angles automatically tuned and spectrum re-measured to confirm its quality before restart. Fig.7 shows the typical measured CB spectra for the vertical and horizontal beam polarizations. The photons polarization was calculated according to a scheme developed in ref. [4,5] and for the central energy bin ($E_\gamma$=1000±20MeV) is equal to $P_\gamma$=0.53 ± 0.02.

The energy calibration of PS-6 channels represents the strict test of the measured CB shape compatibility with one measured by PS-30 pair spectrometer. Fig.8 shows the normalized each to other spectra in the peak region measured simultaneously by PS-30 (full curves) and PS-6 (points). As is seen from figure the spectra are quite similar, no notice able shift is observed in between.

The data on the vertical z-scan at fixed $\Delta z$ = 0 are shown in Fig.4 together with Monte Carlo simulation results. As can be seen in the figure, Monte Carlo predictions satisfactorily describes the experimental data. From Monte Carlo z-dependence Gaussian fit result for the mean value error obtained is app. 0.1 mm. These two tests have established a reachable precision of the geometrical arrangement of a PS – 6 detectors.

As a test of the apparatus systematic uncertainties, coming in particular from the beam, detectors and monitoring system instabilities, the measurements of CB spectra with disoriented crystal (amorphous spectrum) has been carried out and coincidence rate of PS-6 telescopes compared with the average of the rate, obtained for two orthogonal orientations of photons polarization at CB peak energy of 1000 MeV. The obtained difference of those rates, normalized to the number of photons, didn't exceed 2-3 % (stat.), as it was expected.

Unfortunately the tight time schedule of accelerator's work didn't allow to fulfill the test on the vertical slit positioning, without which one may achieve a source of visible change in analyzing power $A_{exp}$, so we have made the preliminary measurements without slit. The measurements were carried out at $\Delta z$=10mm, the half-height of the gap between the counters above the polarimeter PS-6 median plane and those below it, and at CB peak energy 1000 MeV. The asymmetry value $A_{exp}$ = 0.098 ± 0.011(stat.) was obtained. The systematic error in this value comes mainly from the geometrical adjustment uncertainty, that has been evaluated by means of Monte Carlo simulations. The conservative evaluations in the case of polarimeter without slit for the adjustment precision in the x,z plane: $\sigma_x = \sigma_z = 1mm$ corresponds to the $A_{exp}$ uncertainty on the level of app. 0.003.

With this level of systematic uncertainties and increased statistics one may reach a polarization measurement precision in the level of 0.02- 0.03.

Knowledge of $A_{MC}$ asymmetry and calculated photons polarization $P_\gamma$ allow to determine an expected asymmetry using an expression (2). The results are shown in Fig.9 as a function of $\Delta z$ together with the measured value of $A_{exp.}$ As is seen from figure, the values of expected and measured asymmetries are agreed well within the error bars. For comparison the results of two polarimetry methods , one direct and the other indirect on Fig.10 are presented the measured photon beam polarization at $E_\gamma = 1000 \pm 20$ MeV together with calculated curve of polarization by CBSA method [4]. As may be seen the measured value of polarization $P_\gamma^{exp} = 0.565 \pm 0.06$ (stat.) are agreed with calculated.

As it was mentioned in the beginning of the section 3 the measurements of the z-dependence of $A_{exp}$ asymmetry and its comparison to a Monte Carlo predictions is an important test of the polarimeter performance as well as allows an extraction of the atomic form factor over the polarimeter acceptance range. Unfortunately these measurements was not accomplished by the reasons mentioned above.

## 5. Conclusion

Although the time shortness didn't allow to reach a planned precision of CB polarimetry in the level of 0.02-0.03, an experimental method was reliably tested in many details and expected feasibility confirmed. This activity will be continued and we hope to carry out all necessary measurements with full set of the tests, particularly on the correct positioning of the vertical slit which shall noticeable increase the analyzing power of polarimeter by app. 0.25 .

**Acknowledgements**

Authors are indebted to the synchrotron staff and his leader Valery Nikogosyan for their efforts of accelerator running, directorate of YERPHI for the support and funding. The work was also supported by CRDF grant AP2-2305-YE-02, that is acknowledged.

**Figure captions**

Fig.1. Layout of the experimental setup
Fig.2. Sketch of the polarimeter PS-6
Fig.3. The z-dependence of the $A_{MC}$ asymmetry with and without slit (solid and dashed curves respectively), where $z = (z_{up}-z_{down})/2$ is the half-height of the gap in z between the up and down counters. The energy of CB peak is set to $E_\gamma = 1.0$ GeV.

Fig.4. The z-dependence of calculated and experimental yields of ($N_{2up}$ x $N_{2down}$) coincidences at fixed vertical distance between up and down telescopes z = 0.

Fig.5. The x-dependence of $A_{MC}$ asymmetry for the central telescopes ($N_{2up}$ x $N_{2down}$) of PS-6 at fixed z=10mm, where x is defined as a shift of the central telescope relative to its symmetry configuration.

Fig.6. Testing the axial symmetry of PS-6 telescopes' configuration
  (a) an initial position of the telescopes ($N_{2up}$ x $N_{2down}$)
  (b) the final configuration after successive rotations around vertical and horizontal axes

Fig.7. An intensity spectra of PS-30 pair spectrometer measured at CB peak energy setting to $E_\gamma^{peak}$= 1.0 GeV for both vertical and horisantal orientations of the photon beam polarization..

Fig.8. An intensity spectra in a CB peak region ($E_\gamma^{peak}$=1.0 GeV) measured simultaneously by PS-30 and PS-6.

Fig.9. An expected z-dependence of $A_{MC}$ calculated for the central energy bin of PS-6 for the photons polarization $P_\gamma$=0.53±0.02 An experimental result presented is measured at z setting to 10mm.

Fig.10. The measured photon beam polarization at $E_\gamma$= 1000 ± 20 MeV together with calculated curve of polarization by CBSA method [4].

**Reference**

[1] H. Uberall, Phys.Rev., v.103,4, (1956)1055,
    H. Uberall, Phys.Rev., v.107,1, (1957 233
[2] H. Olsen, L.C. Maximon, Phys.Rev. 114 (1959) 887,
    G. Barbielini et al., Phys.Rev.Lett.,v.9 (1962) 396
[3] L.C. Maximon, H.A. Gimm, Phys.Rev. A 23 N1 (1981) 172
    V.F.Boldyshev et al., Phys.Part .Nucl. 25 (1994) 292
[4] S.Darbinyan et al.,NIM A 554 (2005)75
[5] H.Hakobyan et al., YERPHI-908(59)-86 (1986)
    H.Hakobyan , G.Karapetyan, YERPHI-1138 (15)-89 (1989)
[6] F.Adamyan et al., NIM A 546 (2005) 376
[7] L.C. Maximon, H. Olsen, Phys.Rev. 126 (1962) 310
[8] R. Avagyan et al. hep-ex/9908048 v.2 (1999)
[9] J.H.Hubbell et al.,J.Phys.Chem.Ref.Data, v.4, No3 (1975) 471
    J.H.Hubbell et al.,J.Phys.Chem.Ref.Data, v.8, No1 (1979) 69*
[10] R. Avakyan et al. Izvestia Acad.Nauk Arm.SSR, Physics,v.10 (1975) 61
[11] A. Avetisyan. et al., YERPHI-1325(20)-91 (1991)

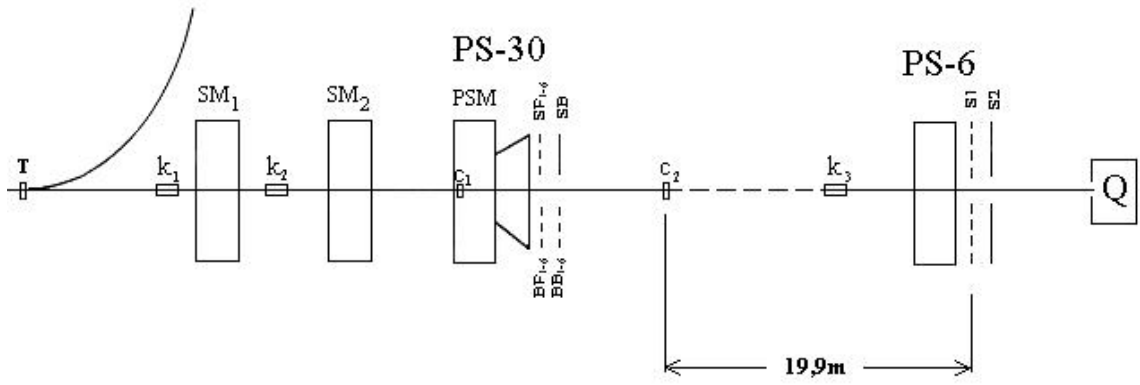

Fig.1

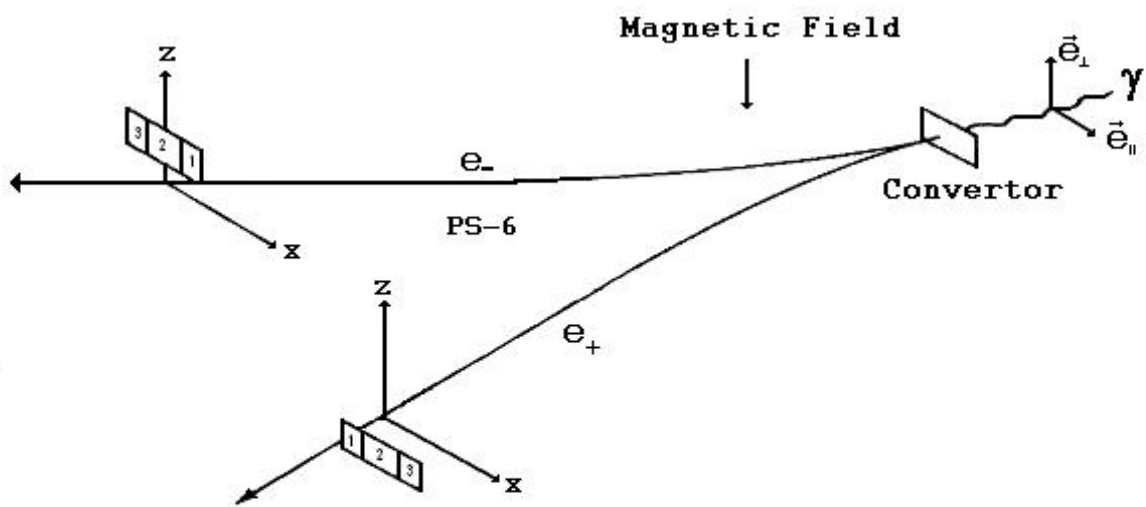

Fig.2

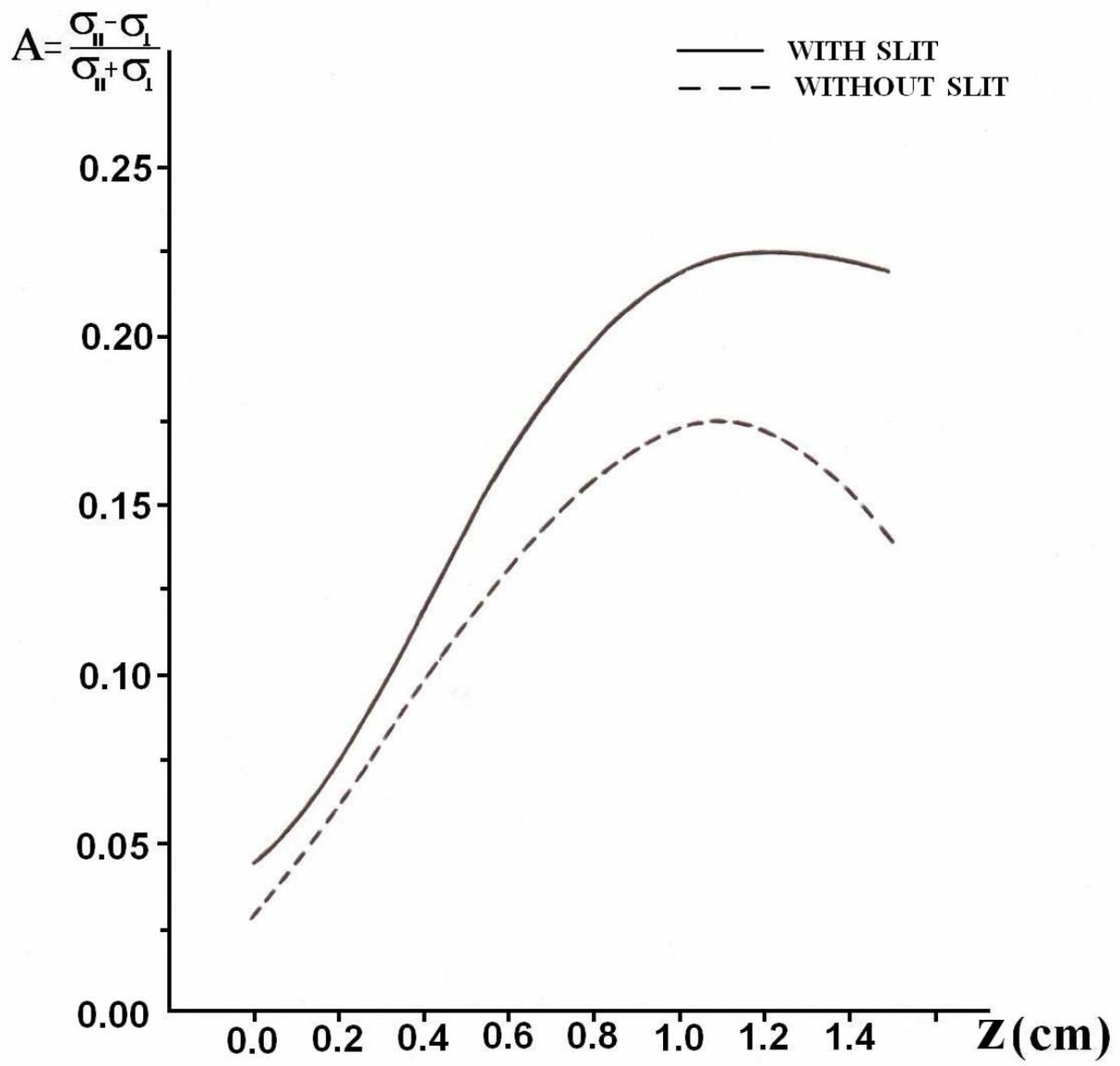

Fig.3

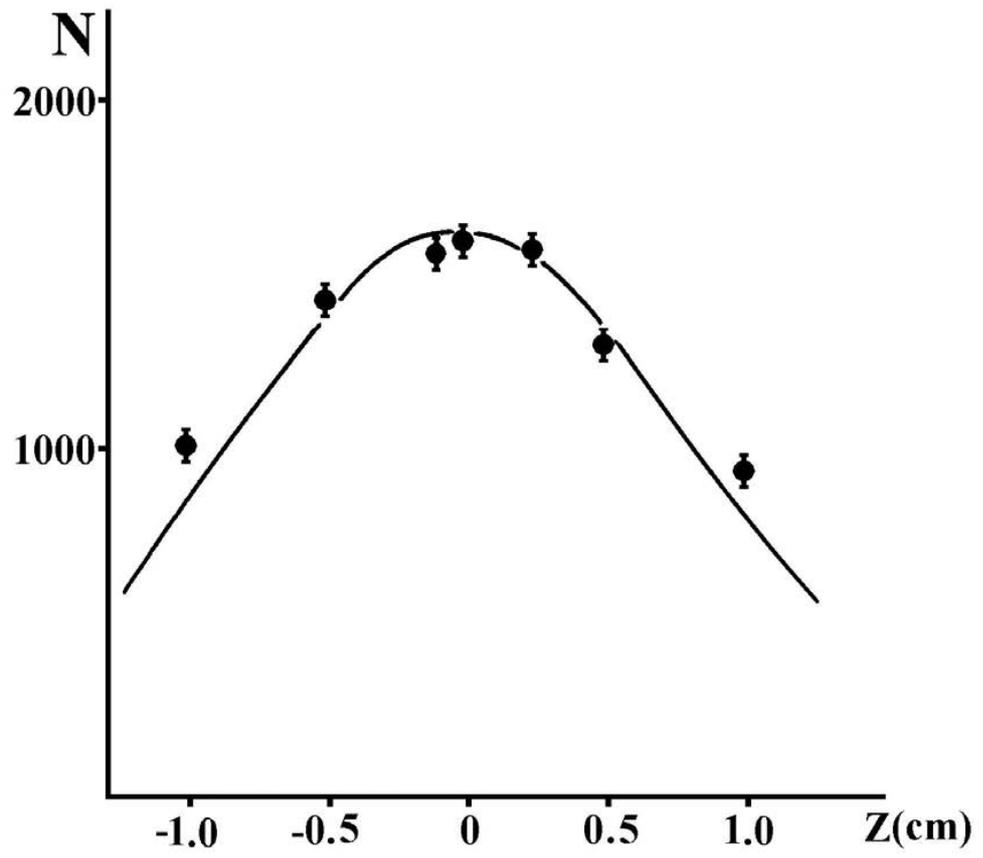

Fig.4

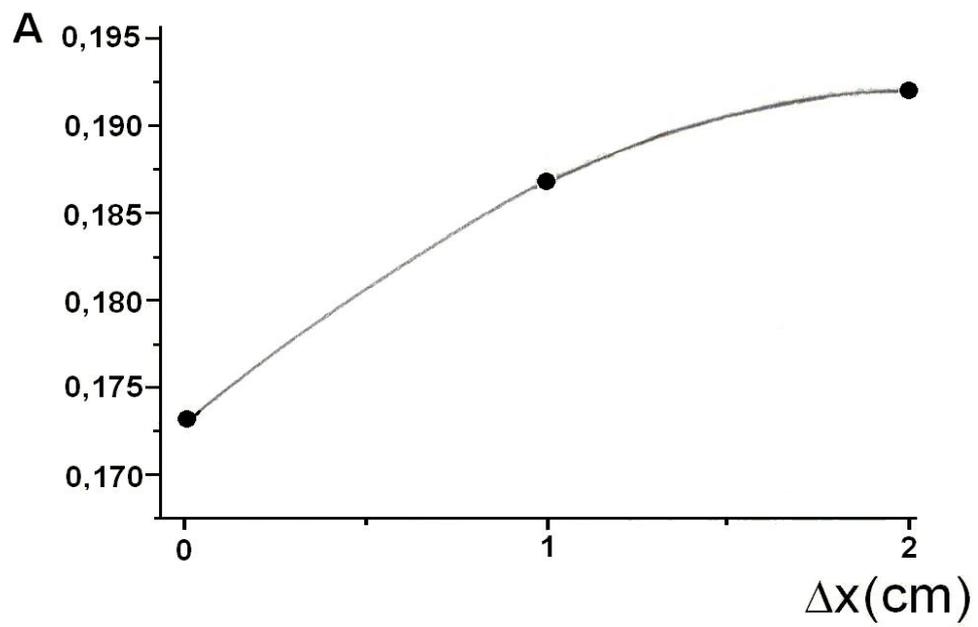

Fig.5

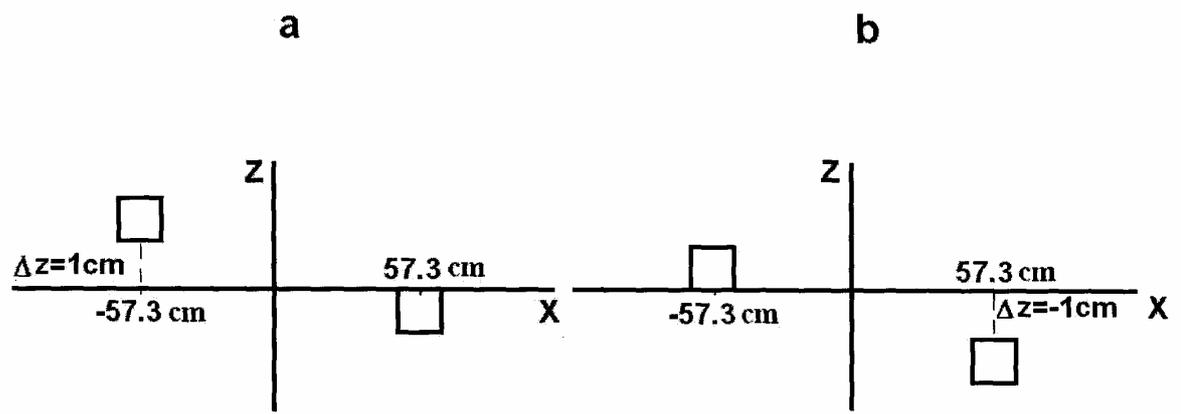

Fig.6

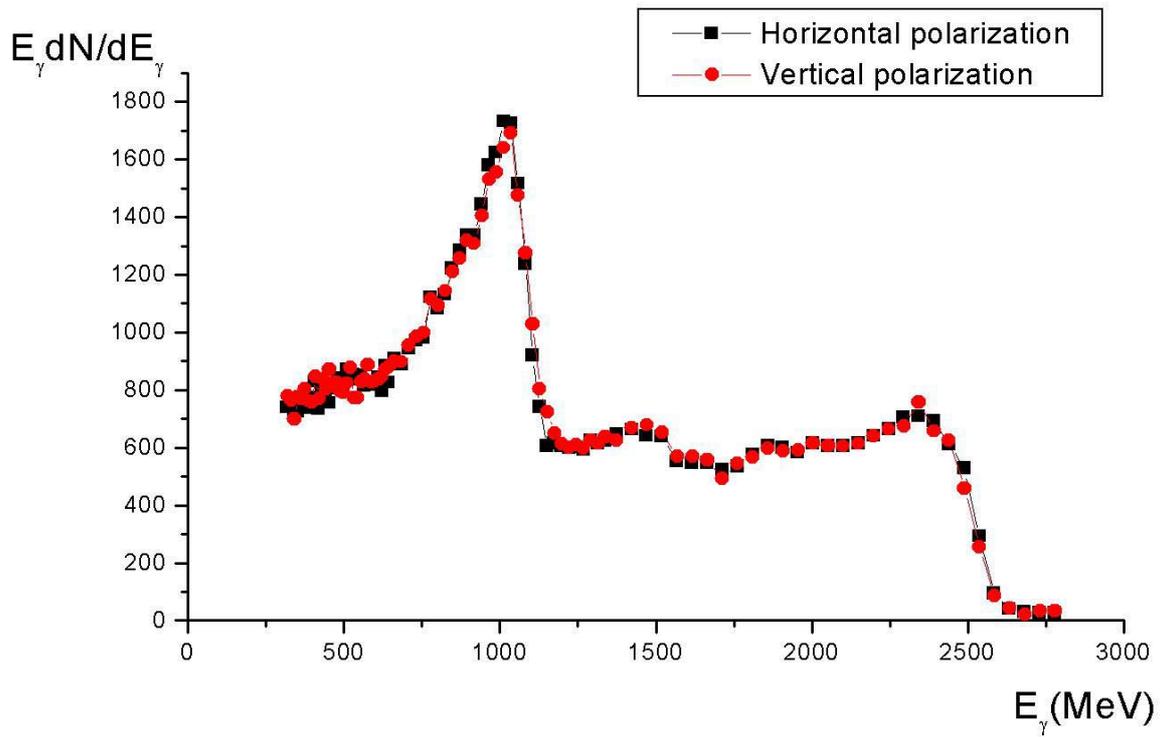

Fig.7

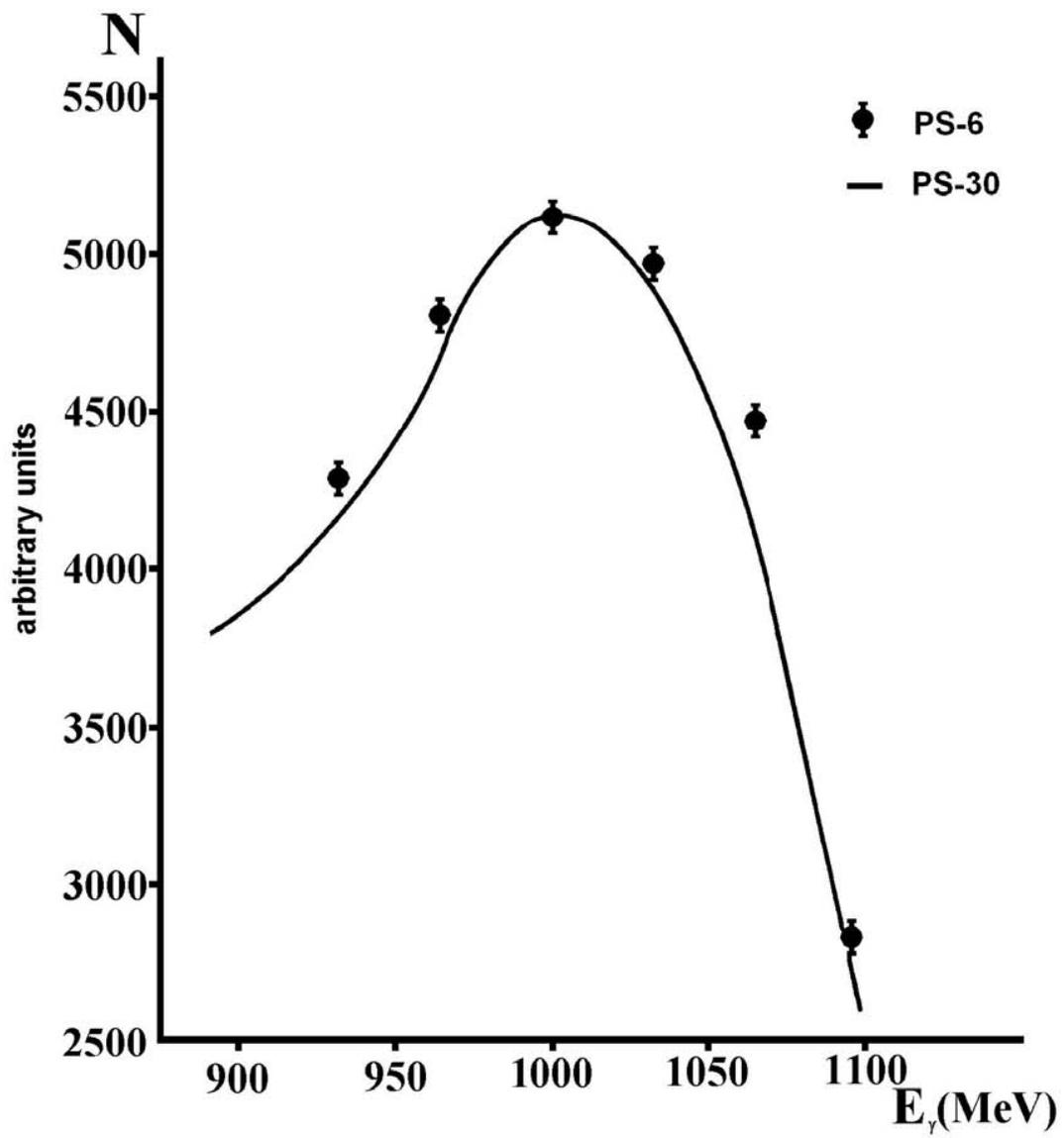

Fig.8

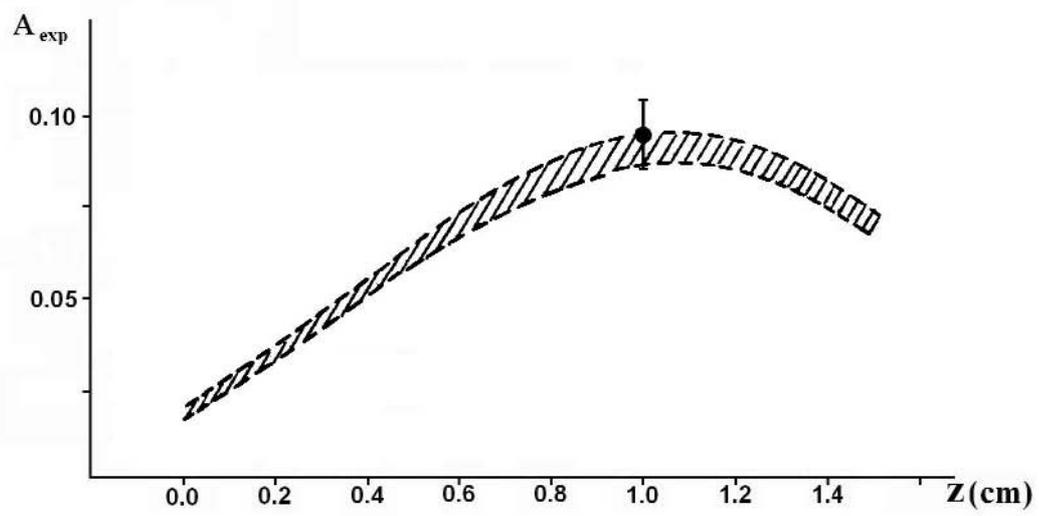

Fig.9

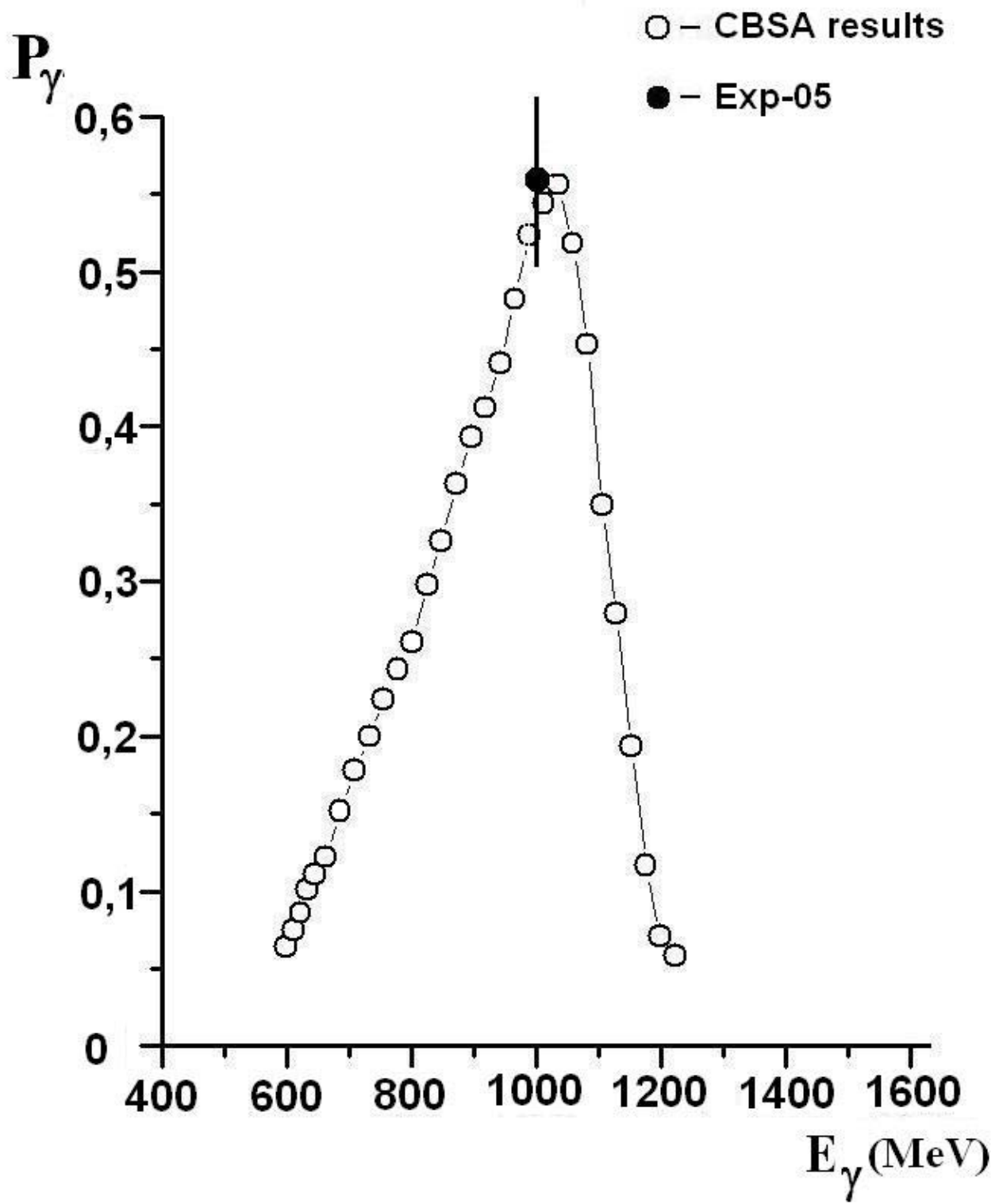

Fig.10